\documentclass[%
 reprint,
 amsmath,amssymb,
 aps,
 hidelinks,
]{revtex4-2}

\usepackage{graphicx}
\usepackage{dcolumn}
\usepackage{float} 
\usepackage{bm}
\usepackage{orcidlink}
\usepackage{tabularx}
\usepackage{booktabs}
\usepackage{natbib}
\begin{document}

\preprint{APS/123-QED}

\title{A triaxial vectorization technique for a single-beam zero-field atomic magnetometer to suppress cross-axis projection error}

\author{R. Dawson\orcidlink{0000-0003-4862-3066}}
\author{M. S. Mrozowski\orcidlink{0000-0001-9616-4058}}
\author{D. Hunter\orcidlink{0000-0003-4177-6027}}
\author{C. O'Dwyer\orcidlink{0000-0003-1922-6504}}
\author{E. Riis\orcidlink{0000-0002-3225-5302}}
\author{P. F. Griffin\orcidlink{0000-0002-0134-7554}}
\author{S. Ingleby\orcidlink{0000-0001-7473-9949}}

\affiliation{%
 Department of Physics, SUPA, University of Strathclyde, Glasgow G4 0NG, United Kingdom
}%

\date{\today}

\begin{abstract}
Zero-field optically pumped magnetometers (OPMs) have emerged as an important technology for biomagnetism due to their ulta-sensitive performance, contained within a non-cryogenic small-scale sensor-head. The compactness of such OPMs is often achieved through simplified detection schemes, which typically provide only single-axis magnetic field information. However, multi-axis static magnetic fields on non-measurement axes cause a systematic error that manifests as amplitude and phase errors across the measurement axis.
Here we present a triaxial operational technique for a compact zero-field OPM which suppresses multi-axis systematic errors through simultaneous measurement and closed-loop active control of the static magnetic fields across all axes. The demonstrated technique 
requires magnetic modulation across two axes while providing static field information for all three axes. We demonstrate this technique on a rubidium laboratory-based zero-field magnetometer, achieving a bandwidth of 380~Hz with sensitivities of $<25$~fT/$\sqrt{\rm{Hz}}$ across both transverse axes and~$65$~fT/$\sqrt{\rm{Hz}}$ along the beam axis. Using the proposed triaxial technique, we demonstrate precise tracking of a 2~Hz triaxial vector test signal and suppression of systematic cross-axis projection errors over an extended period, $\simeq20$~min.

\end{abstract}

\maketitle

\section{Introduction}\label{sec:Intro}
Precise measurement of small, sub-picotesla scale ($<10^{-12}$~T), magnetic fields is crucial in the realm of biomagnetism for both diagnosis and research \cite{Meyer2018,Chen2019cerebral,Zhang2022editorial,Hamedi2022,Hamada1999,Lopes2021}. Optically pumped magnetometers (OPMs), typically in zero-field configurations, present an alternative to traditionally used superconducting quantum interference devices (SQUIDs) for this application. Zero-field OPMs offer similar sensitivity to SQUID magnetometers ($<$~10$^{-{14}}$~T), without the requirement of cryogenic cooling \cite{Dang2010, Allred2002, Ledbetter2008, Fang2014,Brookes2022}. Without cryogens, zero-field OPMs are more easily integrated into small, portable packages \cite{Dawson2023SPIE,westner2021contactless, Brookes2022}
, that facilitate both practical measurements, flexible placement and multi-sensor arrays for spatial mapping of biomagnetic fields \cite{Jas2021,Hill2020,Yang2021}. 
However, in many cases, practical OPMs are realized using compact single-beam optical schemes, resulting in single- or dual-axis magnetic sensitivity. 
Single-axis measurements are susceptible to systematic errors \cite{Borna2022}, and can overlook magnetic field information of a complex field, such as the radial and tangential components of a human heart signal \cite{Watanabe2008, Kwon2010}.

OPMs use laser light to optically pump and probe alkali atoms in vapor phase, detecting the response of the atomic sample to magnetic fields. Zero-field OPMs operate in the spin exchange relaxation free (SERF) regime in which the relaxation effect of spin-exchange collisions is suppressed \cite{Happer1977} providing narrower magnetic resonance linewidths that enhance the overall sensitivity performance. 
In the SERF regime, high rates of ground-state spin-exchange, coupled with slow spin precession, result in long-lived ground-state spin coherence \cite{Ledbetter2008}. Operation in the SERF regime requires a low-field environment, typically provided through magnetic shielding, and a suitably high atomic density 
achieved through heating of the atomic vapor cell. OPMs operating in the SERF regime exploit the ground-state Hanle effect \cite{Castagna2011} which results in a well-defined peak in optical transmission at $|B|$ = 0.

Cross-axis projection error (CAPE) is a term coined to describe the systematic error that arises in zero-field OPMs due to the presence of spurious magnetic fields perpendicular to the sensitive axis of the OPM \cite{Borna2022}. CAPE results in phase errors, time delays, and amplitude errors in recovered magnetic signals, which ultimately reduces the accuracy of source localization in biomedical applications. Methods for minimizing CAPE often include multiple axis field nulling achieved through active feedback and dynamic field compensation \cite{Jia2023,Robinson2022}. In the simplest terms, CAPE can be suppressed through the identification and compensation of the magnetic field across all three axes. 

Three-axis magnetic field extraction schemes are proposed and demonstrated utilizing triaxial modulation \cite{fang2012, tang2021}, multiple laser beams \cite{Seltzer2004,zou2022, Boto2022Quspin}, or single-tone dual-axis modulation \cite{huang2016, Shah2009,Dong2012}. In this paper, we introduce a novel approach: a three-axis single-beam OPM detection scheme utilizing two cross-modulating fields at non-factoring frequencies. This technique derives the beam-axis magnetic field from the second harmonic component of the modulating fields, thus offering a triaxial measurement technique that only requires dual-axis modulation with a single laser beam.

\section{Theory: Single-beam atomic response}
OPMs utilize laser light, tuned to an appropriate atomic transition, to optically pump an atomic ensemble of alkali vapor into a polarized atomic state, resulting in a strong net magnetization. In this polarized state, the atomic ensemble is highly sensitive to changes in magnetic field. The magnetization of the atomic ensemble evolves with respect to the magnetic field, and this evolution is detected by an optical absorption measurement, which may utilize the same laser beam used to optically pump the sample\cite{Shah2009}.

By considering the net magnetisation induced by optical pumping we can derive a semi-classical model that can simulate the atomic response to multi-axis magnetic fields. The Bloch equations provide a foundation to phenomenologically describe atomic magnetization \cite{Bloch1946}. Equation~(\ref{Eq. Bloch Equation for dipole}), describes the evolution of the net magnetization of the atomic ensemble, $\underline{M}$, with respect to time such that;
\begin{equation}
\underline{\dot{M}} = \gamma \underline{M} \times \underline{B}(t) - {\Gamma}\underline{M} + \Gamma_P \underline{M_0}~,\\
\label{Eq. Bloch Equation for dipole}
\end{equation}
where $\gamma$ is the gyromagnetic ratio and $\underline{B}(t)$ is the time-varying magnetic field experienced by the atoms. The relaxation matrix, ${\Gamma}$, consists of the longitudinal relaxation rate, $\Gamma_1$, and the transverse relaxation rate, $\Gamma_2$. For zero-field OPMs in which spin-exchange relaxation is suitably suppressed due to operation in the SERF regime, we can approximate $\Gamma_2 = \Gamma_1 = \Gamma$. $\Gamma_P$ is the relaxation due to the equilibrium magnetization $\underline{M_0}$, due to optical pumping along the beam axis (defined as the $z$-axis for our analysis).

The atomic response under multi-axis modulation is modeled through expansion of the Bloch equations for the system, Eq.~(\ref{Eq. Bloch Equation for dipole}), where along each axis a combination of two types of fields are applied:
\begin{enumerate}
    \item A static field; 
    \begin{align}
        {B_x}_0,{B_y}_0\mbox{ \& }{B_z}_0~. \nonumber
    \end{align}
    \item A modulated magnetic field, ${B_x}_{\rm{Mod}}$ \& ${B_y}_{\rm{Mod}}$, across the transverse axes, $x$- and $y$-axes respectively, where;
    \begin{align}
        {B_x}_{\rm{Mod}}&= {A}_x \sin {(2\pi f}_x t)~,\label{eq_Bx_mod}\\
        {B_y}_{\rm{Mod}}&= {A}_y \sin {(2\pi f}_y t)~,\label{eq_By_mod}
    \end{align}
    with corresponding modulation magnetic field amplitudes, $A$, in the order of 10's of nT, and modulation frequencies, $f$, in the order of 100's of Hz. The modulation index, the ratio of modulation amplitude amd modulation frequency, requires low magnetic field values for optimal linear modulated atomic response \cite{Yin2022}.
\end{enumerate}
As such, the change in magnetization vector with time, for a zero-field single-beam OPM, can be expressed as:

\begin{align}
\dot{\underline{M}} &= \begin{pmatrix}
-\Gamma & \beta_z & -\beta_y\\
-\beta_z & -\Gamma & \beta_x\\
\beta_y & -\beta_x & -\Gamma
\end{pmatrix} \underline{M} +
\begin{pmatrix}
0\\
0 \\
\Gamma_P M_0 \end{pmatrix}\label{Eq: CAPE Bloch}~,
\end{align}
where; 
\begin{align}
    \beta_x &= \gamma({B_x}_0 + {B_x}_{\rm{Mod}}) ~,\label{beta_x}\\
    ~\beta_y &= \gamma({B_y}_0 + {B_y}_{\rm{Mod}})~,\label{beta_y}\\
    \beta_z &= \gamma{B_z}_0~. \label{beta_z} 
\end{align}

For slowly changing magnetic fields, we can set $\dot{\underline{M}} =0$, to obtain an analytical solution for each axis magnetization \cite{Seltzer2004}. Omitting the common factor of $\Gamma_P M_0$, thus; 
\begin{align}
    \frac{M_x}{\Gamma_P M_0} &= \frac{-\Gamma\beta_y + \beta_x\beta_z}{\Gamma(\Gamma^2 + \beta_x^2 + \beta_y^2 + \beta_z^2)}
    ~,\label{Eq: CAPE Mx}
    \\
    \frac{M_y}{\Gamma_P M_0} &= \frac{\Gamma\beta_x + \beta_y\beta_z }{\Gamma(\Gamma^2 + \beta_x^2 + \beta_y^2 + \beta_z^2)}
    ~,\label{Eq: CAPE My}
    \\
    \frac{M_z}{\Gamma_P M_0} &= \frac{\Gamma^2 + \beta_z^2 }{\Gamma(\Gamma^2 + \beta_x^2 + \beta_y^2 + \beta_z^2)}
    ~.\label{Eq: CAPE Mz}
\end{align}

Conventionally, the equations for net magnetization, Eq.~(\ref{Eq: CAPE Mx}-\ref{Eq: CAPE Mz}) can be simplified for small fields \cite{Seltzer2004}. However, in this case, the non-negligible amplitude of the modulating fields results in significant cross terms ($\beta_i\beta_j$). We solve equations \ref{Eq: CAPE Mx} - \ref{Eq: CAPE Mz} using step-wise numerical integration, with a step size of $dt = 10~\mu$s across total time $t = 100$~ms. The equilibrium condition is achieved when the modulated response of the atoms reaches a steady-state sine wave at the modulation frequency. Typically, the atoms take around 10 ms to reach this equilibrium state. Thus using this approach, we calculate the time-evolved response of the net magnetization vector and the detected light transmission. 

Demodulation of the measured signal allows us to analyze changes in the DC level, first-harmonic ($f_x$, $f_y$) and second-harmonic ($2f_x$, $2f_y$) components separately or linear combinations thereof. Combined with the derivation above, this harmonic analysis allows separate determination of ${B_x}_0,{B_y}_0\mbox{ \& }{B_z}_0$. Therefore, real-time measurement of the first and second-order components of the modulation fields, through lock-in amplifiers, allows for the extraction of three-axis vector measurement of any external magnetic field.

\section{Experimental methods}\label{sec:Method}

\begin{figure}[t]
\includegraphics[trim=15 0 0 0 mm, clip=true
]{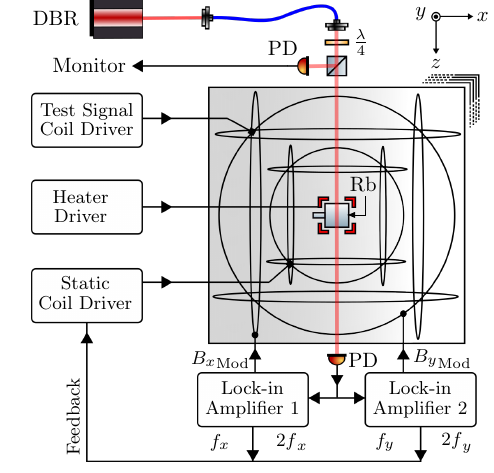}
\caption{\label{fig:setup} Setup of the rubidium zero-field magnetometer. The OPM operates inside a four-layer mu-metal shield (Twinleaf MS-1L). The laser light from the Distributed Bragg Reflector (DBR) is delivered to the atoms using an optical fiber. Multiple sets of three-axis Helmholtz coils added inside the shield provide triaxial static field control, dual-axis modulation and triaxial test signals. The driving and control electronics including Data Acquisition (DAQ) unit, coil driver \cite{Mrozowski2023} and cell heater operate outside the shielding.}
\end{figure}

The setup of the single-beam zero-field OPM is shown in Fig. \ref{fig:setup}. The borosilicate glass atomic vapor cell ($10\times10\times10$~mm), containing enriched rubidium-87 atoms and 200~Torr of nitrogen buffer gas is heated to $\simeq~150~^{\circ}$C, to achieve an atomic density $\rho \simeq 10^{14}$~cm$^{-3}$, by AC resistive heating at 300~kHz. The OPM is housed inside a four-layer mu-metal shield (Twinleaf MS-1L), with a shielding factor of $\simeq~10^6$, to attenuate any ambient magnetic fields and provide a low nT-level magnetic field environment. The expected magnetic Johnson noise at the centre of this shield is around $16$~fT/$\sqrt{\rm{Hz}}$\cite{Twinleaf2024} which is the dominant source of noise in the system. The magnetic Johnson noise could be improved with the introduction of a ferrite layer\cite{Kornack2007}.

The rubidium atoms are pumped along the $z$ direction by a 10~mW Distributed Bragg Reflector (DBR) laser, $\lambda \simeq 795$~nm, $\simeq3$~GHz detuned from the D$_1$ resonance of the rubidium $F = 2 \rightarrow F' = 1$ hyperfine transition, as measured on a reference cell. Use of a spectroscopic reference cell allows for clear identification of the expected optical peaks. The laser light is circularly polarized by a $\frac{\lambda}{4}$ waveplate and split with a 50:50 non-polarising beam splitter, providing two beams of equal power to allow for differential measurement and common-mode noise suppression. The transmission of laser light through the atoms is measured by differential measurement of the picked-off light, through a monitor photodiode, and the remaining light, which is incident through the cell onto a photodiode.

Multiple coils are utilized for operation and demonstration, including a set of three-axis, ($x,y,z$), Helmholtz coils to nullify residual static magnetic fields, using a custom low-noise current driver \cite{Mrozowski2023}. A set of two-axis, ($x,y$), Helmholtz coils, to modulate the magnetic field across the $x$- and $y$-axes. A further set of three-axis coils, ($x,y,z$), mounted inside the innermost shield layer, used for application of magnetic test signals.

\begin{figure}[t]
\centering
\includegraphics[trim=75 -10 0 0 mm, clip=true
]{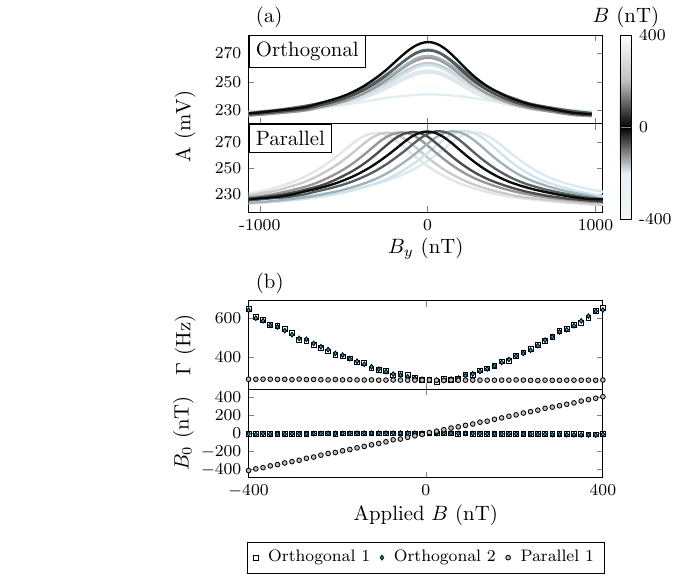}
\caption{Results of coil orthogonality testing, by measuring magnetic resonances whilst applying magnetic fields to other coils that are either orthogonal or parallel to the measurement axis. \textbf{a}) Hanle resonance recovered from sweeping the field, $B_y$, along a transverse axis ($y$-axis). Magnetic fields, $B$, are simultaneously applied (amplitude is indicated by color) across using other coils, either orthogonal ($x$- and $z$- axis) or parallel ($y$-axis), as indicated by the label. \textbf{b}) Extracted $\Gamma$, and $B_0$ for the Hanle resonances measured on the $y$ (parallel) axis under the influence of fields applied on the two orthogonal axes.
Orthogonal fields did not significantly shift the resonance center on the parallel axis.
} 

\label{Fig: orthogonality}
\end{figure}

A high degree of orthogonality between these Helmholtz coils is essential to ensure accurate magnetic vector identification. Orthogonality of the coils is verified through repeated measurement of a single-axis Hanle resonance \cite{Dawson2023} whilst applying magnetic fields on other coils that are either orthogonal or parallel to the measurement axis. Figure.~\ref{Fig: orthogonality}(a) demonstrates how orthogonal fields affect the resonance full-width at half-maximum
, $\Gamma$, and parallel fields shift the resonance zero-field value, $B_0$, determined by the same method as \cite{Dawson2023}. If an applied magnetic field causes a change in both $\Gamma$ and $B_0$, a non-orthogonality is present between these coils. Figure.~\ref{Fig: orthogonality}(b) illustrates the change in $\Gamma$ and $B_0$ for three Helmholtz coils with a high degree of orthogonality. All coils were validated by this method, non-orthogonality limited to $\leq 1\%$ of the applied field.

\begin{figure}[t]
\includegraphics[trim=10 0 0 0 mm, clip=true]{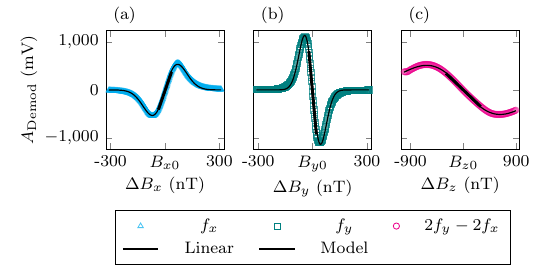}
\caption{Lock-in features for the \textbf{a)} $x-$, \textbf{b)} $y-$ and \textbf{c)} $z-$axes, extracted from the amplitude of the demodulated response, $A_{\rm{Demod}}$, measured with respect to a sweeping magnetic field applied along the corresponding axis. Color indicates the frequency of demodulation which occurs at $f_{x/y}$ for the $x$- and $y$-axes, and at $2f_{y}-2f_{x}$ for the beam axis, $z$. The thick black lines indicate the linear response region and dynamic range. The thin black line shows the modeled response found through a numerical solution to Eq.~(\ref{Eq: CAPE Mx}-\ref{Eq: CAPE Mz}).
${B_x}_0$, ${B_y}_0$ and ${B_z}_0$ indicate the zero-field point for the $x-$, $y-$ and $z-$axis respectively. }\label{fig:lock in features}
\end{figure}

\section{Triaxial lock-in features}\label{sec:lockin}

To operate the zero-field OPM, initially, the residual fields (${B_x}_0$, ${B_y}_0$ \& ${B_z}_0$ for the $x$-, $y$- \& $z$-axes respectively) are detected using 2D and 1D Hanle resonances in the manner described in \cite{Dawson2023}. These fields are canceled using the Helmholtz coils described in Section~\ref{sec:Method}.

In a typical single-axis detection scheme \cite{Dawson2023SPIE} the static field, $B$, on the measurement axis is swept whilst simultaneously applying a modulating field, $B_{\rm{mod}}$ along the same axis. By demodulating the resultant signal, $A_{\rm{Demod}}$, at the modulating field frequency for each value of $B$, a dispersive signal may be obtained.

The triaxial detection scheme is similar, however in this scheme, two modulating fields are simultaneously applied on the transverse, $x$- and $y$-, axes. Dispersive features can be recovered for all three axes by sweeping the field across each axis iteratively, and demodulating the resultant signal, $A_{\rm{Demod}}$, at ${1f}_x$, ${1f}_y$ \& ${2f}_y- {2f}_x$ for the $x$-, $y$- \& $z$-axes respectively, as shown in Fig.~\ref{fig:lock in features}. These dispersive features indicate the response that changing magnetic fields on each axis induce in the transmission signal after lock-in detection. The modeled response found through a numerical solution to Eq.~(\ref{Eq: CAPE Mz}) is also indicated in Fig.~\ref{fig:lock in features}. The experimentally measured Hanle resonance linewidth and amplitude are used to scale the model to volts. The zero-field values, ${B_x}_0$, ${B_y}_0$ and ${B_z}_0$, are also experimentally measured and passed to the model to improve fit.

The linear regions of the dispersive curves, as indicated in Fig.~\ref{fig:lock in features}, define the dynamic range, $R_{dyn}$, of each axis, in which the measured response of the signal is approximately proportional to the applied magnetic field. The linearity within the dynamic range is important for the extraction of the gradient, which provides calibration of the measured atomic response from the voltage from the photodiode to a magnetic field value. 

\begin{figure}[t]
    \centering
    \includegraphics[trim=35 0 0 0 mm, clip=true]
    {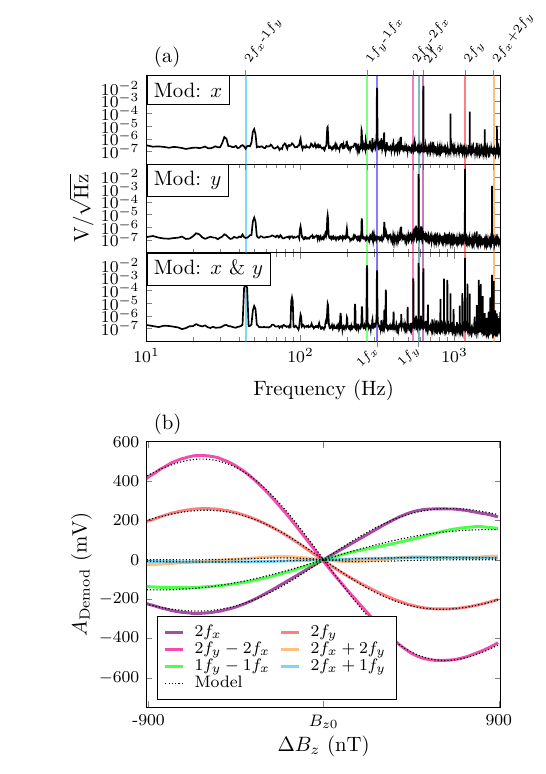}
    \caption{\textbf{a}) Measured magnetic noise floor of the sensor for 3 cases; 1) Mod: $x$, only modulating the $x$-axis magnetic field at the frequency of $1f_x$, 2) Mod: $y$, only modulating the $y$-axis magnetic field at the frequency of $1f_y$, and 3) Mod: $x$ \& $y$, simultaneously modulating the $x$ and $y$-axes magnetic fields at the frequencies of $1f_x$ and $1f_y$ respectively. Various frequencies are indicated through colored lines and labels, which are combinations of the first harmonic, $1f_x$ \& $1f_y$, and second harmonic, $2f_x$ \& $2f_y$, of the applied modulation frequencies. \textbf{b}) The response of the atoms as a function of the applied field, $\Delta B_z$, on the $z-$ axis along the beam axis with respect to the zero-field point, $B_{z_0}$. Various responses are extracted through demodulation of the measured voltage, $A_{\rm{Demod}}$, at specific frequencies, as indicated through colors. The dotted line indicates the modeled response found through a numerical solution to Eq.~(\ref{Eq: CAPE Mx}-\ref{Eq: CAPE Mz}), scaled by the measured Hanle resonance linewidth and amplitude.
    }
    \label{Fig: 2f investigation}
\end{figure}

\begin{figure}
\centering
\includegraphics[trim=70 0 0 0 mm, clip=true
]{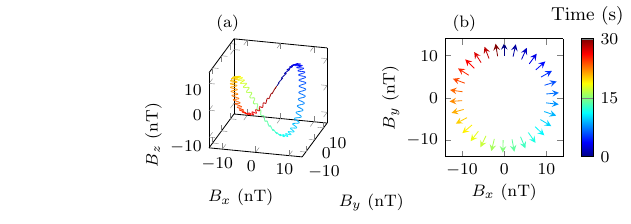}
\caption{Rotating 1~nT amplitude 2~Hz modulating test signal in 3D space, with respect to time (colored). \textbf{a}) Full test signal in all three axes. \textbf{b}) Arrows indicate the vector direction of the 2~Hz modulating test signal across the $xy$-plane.}\label{Fig: Helix test}
\end{figure}

Extraction of the $z$-axis dispersive feature, Fig.~\ref{fig:lock in features}(c), from the second-harmonic component of the transverse modulation field allows for full vector field measurement, which can be used for full field compensation suppressing CAPE systematic errors.

\begin{figure*}
\includegraphics[trim=60 0 0 0 mm, clip=true]{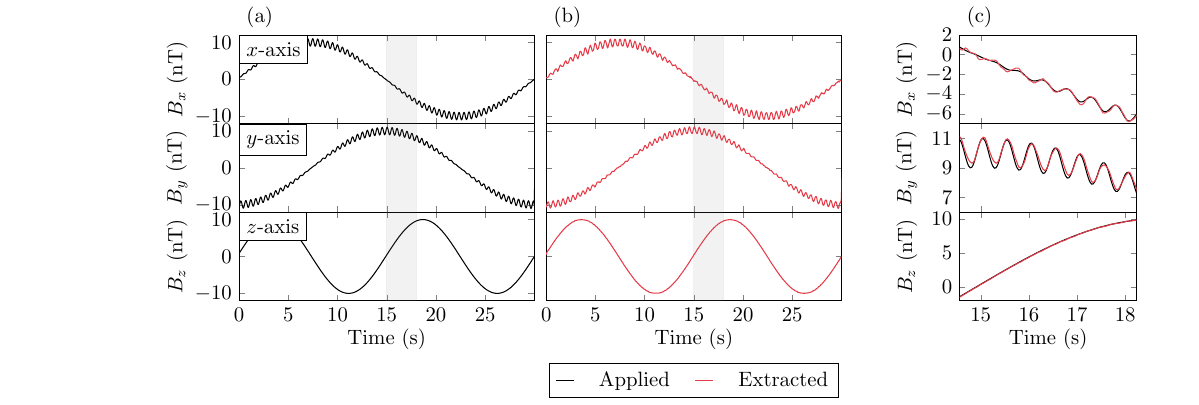}
\caption{Application of a 2~Hz sine test  signal, rotating throughout the $xyz$-space. The color indicates if the field is applied or extracted. \textbf{a)} Each axis component of the applied magnetic field. \textbf{b)}  Each axis component of the recovered magnetic field extracted from each axis PID in closed-loop running mode. \textbf{c)} A 3-second section of the same applied and extracted magnetic fields.
}\label{fig:triaxial disco pringle}
\end{figure*}

In Fig~\ref{Fig: 2f investigation}(a) we explore the effect of multi-axis modulation by analyzing the square root of the power spectral density. Here, we see the effect of modulating in the $x$-axis only and the $y$-axis only, where signal peaks can be seen at the corresponding modulation frequency. We also show the effect of modulating the $x$- and $y$-axes simultaneously, producing signal peaks correlating to the beat of the two frequencies, at various combinations of the two modulating fields as indicated, including ${2f}_x-{1f}_y$, ${2f}_y-{2f}_x$ and ${2f}_x +{2f}_y$ as indicated in Fig~\ref{Fig: 2f investigation}(a).

To empirically ascertain which frequency reference produces the dispersive with the largest gradient, and subsequently optimum sensitivity, we measured the demodulated response of the atoms as a function of the applied field, $\Delta B_z$, on the $z-$ axis at various combinations of the $1f$ and $2f$ values of the modulating fields, shown in Fig~\ref{Fig: 2f investigation}(b). Here we can see the differential of the second harmonic of both modulating fields, ${2f}_y-{2f}_x$, produces the optimal response. These responses correlates to the simulated response found through a numerical solution to Eq.~(\ref{Eq: CAPE Bloch}) indicated by the dotted line, where the measured Hanle resonance linewidth (nT) and amplitude (V) are used to scale the model to volts.

\section{Triaxial operation}

To utilize the established triaxial dispersive features, Fig.~\ref{fig:lock in features}, we use software lock-in amplifiers to detect deviation from zero-field in each axis, and actively correct for this deviation through the application of the corresponding correcting field using a static field coil on the relevant axis. For each axis, a software proportional-integral-derivative (PID) controller is used to calculate the correcting static field value, using the extracted gradient, mV/nT, to convert the measured voltage to field. Thus, through closed-loop feedback and control, we return the atoms to a zero-field environment.

To test the performance of the closed-loop operation, we apply a 2~Hz test signal that is rotating triaxially, as seen in Fig~\ref{Fig: Helix test}(a). The vector direction of this rotating field from the $xy$-plane is shown in Fig~\ref{Fig: Helix test}(b). The test signal consists of three oscillations, one per axis, simultaneously. This allows to test accurate reconstruction of the 2~Hz signal in the presence of 3 component offsets. This aims to mimic a slow oscillating biomagnetic field, such as a heartbeat signal, in the presence of magnetic field changes within magnetic shielding or magnetically shielded room.

The atomic response under the multi-axis static field is seen in Fig.~\ref{fig:triaxial disco pringle}.
Fig.~\ref{fig:triaxial disco pringle}(a) shows each axis component of the applied magnetic field (defined in Fig~\ref{Fig: Helix test}), and  Fig.~\ref{fig:triaxial disco pringle}(b) shows the recovered magnetic field along each axis extracted from the PID in closed-loop running mode. Figure.~\ref{fig:triaxial disco pringle}(c) highlights a 3-second section of the applied and extracted signals, to highlight the correlation between these signals. The calculated RMS error between the applied and extracted field, scaled to the test signal amplitude, is $<0.3$~\% across each axis. Here, we have demonstrated clear extraction of a triaxial magnetic field, through modulation in only two axes, using triaxial lock-in detection and PID closed-loop control.

\section{Suppression of CAPE}
Closed-loop triaxial lock-in operation allows for the extraction of multi-axis magnetic fields, as shown in Fig.~\ref{fig:triaxial disco pringle}, however, operation in this scheme also allows for active cancellation of any direction magnetic fields. Therefore, through active maintenance of zero-field across the atoms in all axes, we can effectively suppress the phase and amplitude errors caused by CAPE to ultimately increase the accuracy of source localization in biomedical applications.

\begin{figure}
\centering
\includegraphics[trim=58 0 0 0 mm, clip=true 
]{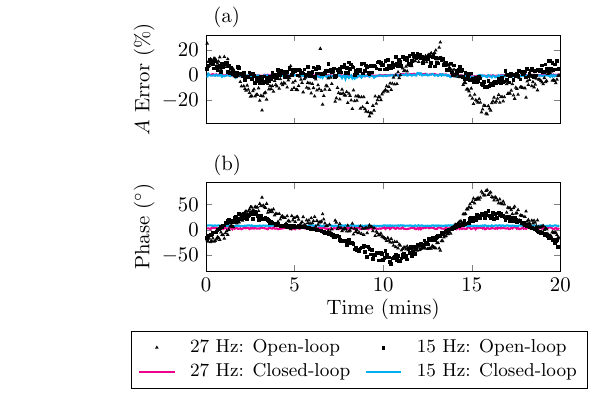}
\caption{Suppression of CAPE-induced measurement artefacts due to off-axis magnetic noise observed in the apparent amplitude and phase of on-axis test signals. \textbf{a}) Sensor signal relative amplitude ($A$) error observed in the presence of off-axis magnetic noise (2~mHz sinusoidal $B_y$ and a 3~mHz ramp $B_z$) for 15~Hz and 27~Hz sinusoidal $B_x$ test signals (black dots and triangles, respectively). In triaxial closed-loop operation, amplitude error of $<2$~\% is observed for both 15~Hz and 27~Hz sinusoidal $B_x$ test signals (blue and pink lines, respectively). \textbf{b}) Sensor signal phase response under the same conditions, showing consistent phase response at both frequencies under triaxial closed-loop operation. 
\label{Fig: Phase and amplitude errors}}
\end{figure}

To demonstrate long-term suppression of CAPE, we investigate the recovery of the amplitude and phase of a 1~nT sine test signal, applied along the $x$-axis, shown in Fig.~\ref{Fig: Phase and amplitude errors}, across a long time frame, $>20$~minutes. To mimic a magnetically noisy environment, we apply slowly oscillating fields, 2~mHz sine and 3~mHz ramp, across the $y$- and $z$- axes respectively.

To extract the amplitude and phase of the measured response at the test frequency, we window the power spectral density response into 2-second intervals where the phase of demodulation is selected to maximize the test signal amplitude, to account for delays between the application of the test signal and atomic response. Amplitude error is calculated from the measured amplitude with respect to the known applied amplitude (1~nT). We used this method to test two frequencies of the test signal, 15 and 27~Hz, that represent commonly measured biomagnetic signals. 

Fig.~\ref{Fig: Phase and amplitude errors} shows the measurement for each test signal, in closed-loop operation mode (colored lines) and without any closed-loop feedback (points) referred to as open-loop mode, as indicated. We can see that in open-loop mode the amplitude errors, Fig.~\ref{Fig: Phase and amplitude errors}(a), and the extracted phase, Fig.~\ref{Fig: Phase and amplitude errors}(b), vary wildly across the measurement time. In closed-loop operation mode, the amplitude error and phase deviation have been almost entirely suppressed, effectively suppressing CAPE.

\begin{figure}[b]
\centering
\includegraphics[trim=20 -10 0 0 mm, clip=true
]{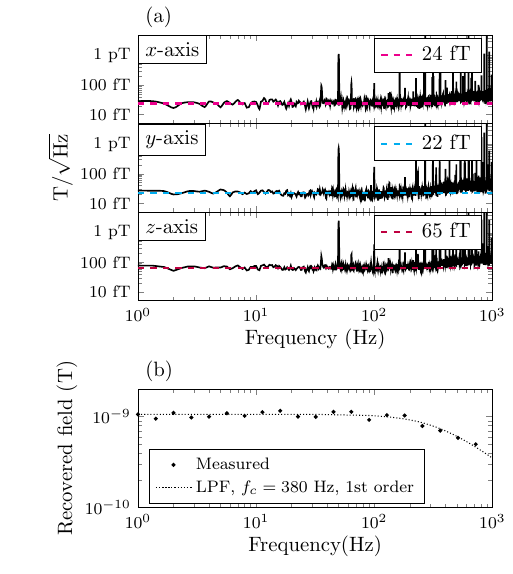}
\caption{\textbf{a}) The measured magnetic noise floor of the sensor for each axis is labeled and scaled by bandwidth, the dashed line indicates the geometric mean calculated in the frequency band of interest (2-100~Hz). \textbf{b}) Measured bandwidth extracted through the recovery of the amplitude of an applied 1~nT sine wave applied along the $x$-axis at various frequencies. The dashed line indicates a low-pass filter fit to the measured data, corner frequency, $f_c$, and order are indicated. 
}\label{Fig: bandwidth}
\end{figure}

\begin{table*}
\caption{Summary of the performance of each sensing axis, including modulation amplitude, $A$ and modulation frequency, $f_{{\rm}}$, for the transverse axes. The linear response within the demodulated signal for each axis is used to extract the slope (mV/nT), where the dynamic range, $R_{dyn}$, defines the boundaries of the linear response. The sensitivity along each axis is directly measured by scaling the square root of the power spectral density of each axis by the slope and averaging across the 5 - 20 Hz frequency band.\label{tab2}}
\begin{ruledtabular}
\newcolumntype{C}{>{\centering\arraybackslash}X}
\begin{tabularx}{\textwidth}{cccccc}
Axis & $A$ (nT)  & $f_{{\rm}}$ (Hz) & $R_{dyn}$  (nT) & Slope (mV/nT) & Sensitivity (fT/$\sqrt{\mbox{Hz}}$) \\
\hline
$x$ & 39   & 314   & $\pm$~35      & 12.1     & 24 \\
$y$ & 45   & 548   &  $\pm$~20     & 48.2     & 22 \\
$z$ & -     & -  &  $\pm$~280    &  2.23     & 65\\
\end{tabularx}
\end{ruledtabular}
\end{table*}
\section{Triaxial sensitivity and bandwidth}
To verify the practical use of the proposed triaxial technique, we measure and compare the OPM performance, as indicated by sensitivity and bandwidth, to single-axis operation.

The OPM sensitivity for each axis is extracted from the square root of the power spectral density and scaled by the gradient of the axis-specific dispersive feature, as discussed in Section~\ref{sec:lockin}. The geometric mean is taken of the noise floor across the frequency band of interest, 2~to~100~Hz, to provide a figure of merit for sensitivity. The extracted sensitivity of the zero-field rubidium OPM in single-axis mode is found as ~$19$~fT/$\sqrt{\rm{Hz}}$, for the $y$-axis. The noise floor and extracted sensitivity, whilst running the triaxial technique with dual-axis modulation, for each axis is shown in Fig. \ref{Fig: bandwidth}(a). Here, both transverse axes are sensitive to $<25$~fT/$\sqrt{\rm{Hz}}$, the small degradation of sensitivity is likely due to the additional second axis modulation inducing a small amount of magnetic field noise. Furthermore, the $z$-axes is sensitive to ~$65$~fT/$\sqrt{\rm{Hz}}$, due to a smaller extracted gradient scaling factor, as seen in Table~\ref{tab2}.

The bandwidth of the OPM is measured by recovering the amplitude of a known amplitude sine wave, (1~nT) along the $y$-axis. We repeated this measurement for a range of frequencies, shown in Fig.~\ref{Fig: bandwidth}(b). The dashed line, in Fig.~\ref{Fig: bandwidth}(b), indicate a low-pass filter response fit to the measured data, with a -3~dB corner frequency of $\simeq380$~Hz.

Table~\ref{tab2} summarises the performance of each axis, including measured sensitivity, the relevant modulation amplitude, $A$, and frequency, $f$. From the dispersive lineshapes for each axis, we can exact the gradient to convert the signal to magnetic field, and the dynamic range, $R_{dyn}$, in which the response to a magnetic field is approximately linear.

While the sensitivity for the transverse axes is comparable ($<25$~fT/$\sqrt{\rm{Hz}}$), the gradient scaling factors differ by a factor of 4, as shown in Table~\ref{tab2}. This difference arises due to the modulation amplitude and frequency selected for each axis, which significantly affect both the dispersive shape and gradient. We have observed that by swapping the modulation parameters between the transverse axes, this factor of 4 difference can be reversed, indicating a strong dependence on these modulation parameters. We have previously demonstrated a framework for optimisation in a complex parameter landscape \cite{Dawson2023}.

The similarity between the transverse axes sensitivities despite a factor of 4 gradient different  suggests we are limited by dominant external noise sources. In this system, we are limited by the shield magnetic Johnson noise, $\approx16$~fT/$\sqrt{\rm{Hz}}$~\cite{Twinleaf2024} and photon shot noise for two photodiodes, $\approx9$~fT/$\sqrt{\rm{Hz}}$~\cite{MarcinThesis}, which in quadrature contribute $\approx20$~fT/$\sqrt{\rm{Hz}}$ noise.

 \section{Discussion}\label{sec:Discussion}

In summary, this article provides the analysis and demonstration of a novel technique for extracting triaxial magnetic field information for a single-beam zero-field OPM, including the theoretical framework and practical implementation of the technique. 

We have demonstrated for a single-beam zero-field magnetometer, magnetic modulation of the two transverse axes allows for the extraction of the longitudinal magnetic field information through demodulation of the second harmonic ($2f$) components of the transverse modulations. Demodulation of the $1f$ and $2f$ components of the two modulating fields produces a distinct zero-crossing dispersive feature in response to a change in the magnetic field in each axis. The gradient of the dispersive feature allows for the extraction of signal (Volts) to field (nT) conversion of the measured demodulated signal for each axis. 

We have demonstrated active correction for deviation from zero, across the measured demodulated signal for each axis, which can be used in a closed-loop feedback system using PID controllers for three axes, using only dual-axes modulation. Through closed-loop feedback of all triaxial components, the systematic error caused by cross-axis projection is effectively nulled.

Implementation of the demonstrated triaxial technique, from an on-sensor hardware perspective, only requires one additional modulation coil, in comparison to conventional single-axis SERF magnetometers. Furthermore, the implementation of the triaxial technique does not significantly degrade the sensitivity or bandwidth of the primary measurement axis.

In conclusion, the triaxial technique is well-suited to implementation in a portable sensor of the type demonstrated in \cite{Dawson2024SPIE}, due to only requiring magnetic modulation and feedback, with immediate potential for impact in the key biomedical imaging applications described above.

\begin{acknowledgments}
This research was funded by UKRI grant number EP/T001046/1.
\end{acknowledgments}

\subsection* {Data Availability Statement} 
All data created during this research is openly available from Pure Data. \\
DOI: https://doi.org/10.15129/28a01257-bb56-49fe-a37e-0e088a363894

\nocite{*}

\bibliography{refs}

\end{document}